\documentclass[aps,prl,twocolumn,showpacs,superscriptaddress]{revtex4}
\usepackage{amssymb,amsmath,amsfonts,bm,bbold,amssymb,amsbsy,latexsym, dsfont,array,layout,mathrsfs,color}
\usepackage[normalem]{ulem}

\newcommand{\ket}[1]{\left|#1\right\rangle}

\newcommand{\one}{\mbox{$1 \hspace{-1.0mm}  {\bf l}$}}

\newcommand\C{\hbox{$\mit I$\kern-.7em$\mit C$}}
\newcommand\R{\hbox{$\mit I$\kern-.6em$\mit R$}}

\begin{document}

\title{Quantum frameness for Charge-Parity-Time inversion symmetry}

\author{Michael Skotiniotis}
\email[]{michael.skotiniotis@uibk.ac.at}
\affiliation{Institute for Quantum Science and Technology, University of Calgary, Calgary, Alberta T2N 1N4, Canada}
\affiliation{Institut f\"ur Theoretische Physik, Universit\"at Innsbruck, Technikerstr.\ 25, A-6020 Innsbruck, Austria}
\author{Borzu Toloui}
\affiliation{Institute for Quantum Science and Technology, University of Calgary, Calgary, Alberta T2N 1N4, Canada}
\affiliation{Department of Physics, Haverford College, 370 Lancaster Avenue, Haverford, PA 19041, United States}
\author{Ian T. Durham}
\email[]{idurham@anselm.edu}
\affiliation{Department of Physics, Saint Anselm College, Manchester, NH 03102}
\author{Barry C. Sanders}
\affiliation{Institute for Quantum Science and Technology, University of Calgary, Calgary, Alberta T2N 1N4, Canada}
\date{\today}
 
\begin{abstract}
We develop a theory of charge-parity-time (CPT) frameness resources to circumvent CPT-superselection.
We construct and quantify such resources for spin~$0$, $\frac{1}{2}$, $1$,
and Majorana particles 
and show that quantum information processing is possible even with CPT superselection. 
Our method employs a unitary representation of  CPT inversion by considering the aggregate action of CPT rather than the composition of separate C, P and T operations,
as some of these operations involve problematic anti-unitary representations.
\end{abstract} 


\maketitle

Superselection rules such as charge~\cite{Wick:1952ys,Aharonov:1967zr},
orientation~\cite{BRS03},
chirality~\cite{Wick:1952ys,Collins:2005uq,Gisin:2004fk},
and phase~\cite{Aharonov:1967zr,Rudolph:2001uq,Sanders:2003kx} prohibit certain coherent quantum superpositions and are formally equivalent to the lack of a requisite classical frame of reference~\cite{Bartlett:2007fk}. Superselection rules can be circumvented by consuming appropriate {\it frameness resources}, namely quantum systems whose states are asymmetric with respect to a group~$G$ of transformations associated with the requisite frame of reference~\cite{Gour:2008vn}.  
Here we develop the superselection rule for charge-parity-time (CPT) invariance~\cite{Schwinger:1951kx,Luders:1954fk,Pauli:1955uq}
and construct the corresponding frameness resources for  spins $s=0$, $s=\frac{1}{2}$, $s=1$
as well as for Majorana particles.
We also suggest a procedure whereby such resources can be constructed for higher-order spins. 

Constructing the CPT operator in the seemingly natural way by composing the separate
C, P, and T operations involves undesirable anti-unitary projective representations.
If~$CPT$ were an anti-unitary projective representation, two phase terms $\pm 1$ arise and
cannot be simply eliminated thereby resulting in a doubling of the representation~\cite{Scu04}.
Perfunctory use of the anti-unitary projective representation
unacceptably allows frameness resources to be converted to non-resources under 
symmetry-respecting evolution,
viz., the Hamiltonian commutes with every element of the representation of the group~\cite{Gour:2009ly}.  

Therefore, we construct $CPT$ as an indecomposable unitary projective representation such that $CPT^2=\one$ with~$\one$ the identity transformation,
and the global phase can be removed by defining the operator appropriately.
We apply this approach to the distinct cases of integer and half-odd-integer $s$
and construct the relevant projective unitary representation for CPT as well as the resource states required to lift CPT superselection.
In addition, our strategy allows for the identification of CPT-invariant subspaces capable of storing and transmitting information even in the presence of CPT superselection. 

In the Feynman-Stueckelberg interpretation~\cite{Feynman:1949uq} the image of a particle with mass $m$, spin $s$, linear three-momentum~$\bm{p}$, and energy $E=\sqrt{|\bm{p}|^2c^2+(mc^2)^2}$ under the action of $CPT$ is an anti-particle of the same mass and energy with its spin and three-momentum reversed, and its internal degrees of freedom, such as electric charge, baryon number, and lepton number, inverted.
%leaving~$m$ and~$E$ unchanged.
Employing only the \emph{universally conserved} internal symmetries, we define
the total internal quantum number
\begin{equation}
\label{eq:u}
	u := Q + (B-L)
\end{equation}
for~$Q$ the total electric charge and $B-L$ the difference between total baryon number~$B$ and total lepton number~$L$.
(In some theories $B$ and $L$ are not individually conserved, but $B-L$ is; this is known as the chiral anomaly~\cite{Bell:1969uq}).

As~$m$ and~$E$ are CPT-invariant,
we denote the state corresponding to a particle with total internal quantum number $u$, spin $s$, and linear three-momentum~$\bm{p}$ as $\ket{u,s,\bm p}$.
The state of the corresponding anti-particle with the same mass and energy is 
\begin{equation}
	CPT\ket{u,s,\bm{p}}=\text{e}^{\text{i}\theta_\text{CPT}}\ket{-u,-s,-\bm{p}}
\label{eq:CPTstate}
\end{equation}
for $\theta_\text{CPT}\in[0,2\pi)$ an unimportant global phase.
The state~$\ket{u,s,\bm{p}}$ is technically not normalizable for an infinite region
with continuous~$\bm{p}$
but is well-defined as a distribution in the dual~$\Phi^*$
to the nuclear space of test functions~$\Phi\subseteq\mathscr{H}$ with~$\mathscr H$
a Hilbert space and $(\Phi,\mathscr{H},\Phi^*)$ the Gel'fand triple~\cite{GelfandVilenkin64}, also known as a rigged Hilbert space.
Observables are complex functionals of 
test functions and distributions like~$\ket{u,s,\pm\bm{p}}$.
In our notation, Dirac ``bras'' refer to test functions and Dirac ``kets'' refer to distributions.
Here we employ the Dirac adjoint representation to ensure
covariance and unitarity throughout~\cite{G00}. 

For reference-frame-establishment protocols,
we consider two parties, Alice and Bob,
who can occupy different space-time regions and, moreover, can be moving relative to each other.
Thus Alice's state~$\ket{u,s,\bm{p}}$
is equivalent to Bob's state only up to a Poincar\'{e} transformation~$\Lambda$
that is known by Alice and Bob
plus either a $CPT$ transformation or else a~$\one$ transformation.

Whether Alice and Bob are related by $CPT$ or by $\one$ is unknown to Alice and Bob,
hence the reference-frame problem.
As Alice and Bob know the transformation~$\Lambda$,
Bob can compensate for its effect on quantum information sent to him from Alice
by employing a suitable device that simulates~$\Lambda^{-1}$ after receiving the particle.
Specifically Bob would apply known rotations and boosts as necessary to recover Alice's basis
modulo whether $CPT$ or~$\one$ should also be applied.

For proper operation of Bob's $\Lambda$ compensator,
we must consider that, although rotation generators commute with $CPT$,
boost generators do not.
Therefore, the order in which  Bob compensates for~$\Lambda$ and~$CPT$ matters.
We assume that Bob compensates for the effects of~$\Lambda$ first.
Thus, upon receiving Alice's particle, which is prepared in some (test-function) state
\begin{equation}
\label{eq:state}
	\ket{\psi}=\sum_{u,s}\int\mathrm{d\bm p} \psi(u,s,\bm p)\ket{u,s,\bm p}
\end{equation}
and hence arrives at Bob's location in the state
$\Lambda\ket{\psi}$,
Bob's device effects $\Lambda^{-1}$ in his frame and recovers the original state up to a CPT transformation.  Henceforth, we focus only on the superselection rule pertaining to $\{\one,\, CPT\}$.

To construct unitary projective representations of $\{\one,\, CPT\}$,
we first complete the intermediate step of constructing the set of operators
$\{\one,\,C,\,PT,\,CPT\}$,
which, under composition, form the (abelian) Klein four-group $\mathbb{Z}_2\times\mathbb{Z}_2$.
Then we reduce this group to the subgroup $\{\one, CPT\}$,
which is equivalent to~$\mathbb{Z}_2$.
The representations are constructed with respect to the states~$\ket{u,s,\bm{p}}$
that span the space of distributions~$\Phi^*$.
The resultant orthonormal basis for given labels~$u$, $s$ and $\bm{p}$ is
\begin{align}
\label{eq:subbasis}
	\{\ket{u,s,\bm{p}},
	\text{e}^{\text{i}\theta_\text{PT}}\ket{u,-s,-\bm{p}}=:&\;PT \ket{u,s,\bm{p}},\nonumber\\
	\text{e}^{\text{i}\theta_\text{C}}\ket{-u,s,\bm{p}}=:&\;C \ket{u,s,\bm{p}},\nonumber\\
	\text{e}^{\text{i}\theta_\text{CPT}}\ket{-u,-s,-\bm{p}}=:&\;CPT \ket{u,s,\bm{p}}\}
\end{align}
with $\theta_\text{CPT}=\theta_\text{PT}+\theta_\text{C}$.
and ``$:=$'' and ``$=:$'' notation indicating that the side with the~``$:$''
is defined by the other side.

By restricting to positive~$u$ and positive~$s$ and `forward'~$\bm{p}$
(i.e., $\bm{p}$ restricted to a half-space of~$\mathbb{R}^3$ with respect to some 
defined `forward' direction vector),
the corresponding bases~(\ref{eq:subbasis}) are mutually orthogonal
with the proviso that the continuous nature of~$\bm{p}$ means that
this orthogonality is of the Dirac-$\delta$ form rather than of the Kronecker-$\delta$ form.
The special case $\bm{p}=\bm{0}$ case is of zero measure in the set of all such sub-bases,
hence does not require special treatment.

We now apply our strategy to special cases of particles, namely relativistic particles with $s=0$, $s=\frac{1}{2}$, and $s=1$.
These cases cover almost all particles of interest in physics.
Subsequent to these cases we explain how to extend the results to $s>1$
by using Bargmann-Wigner equations~\cite{BW48}.

Consider a single, massive spin-$0$ particle with total internal quantum number~$u$ satisfying the Klein-Gordon equation
\begin{equation}
	\left(\square-\frac{m^2c^2}{\hbar^2}\right)\psi=0
\label{eq3}
\end{equation}
with~$\square$ the D'Alembertian differential operator~\cite{G00}. The Klein-Gordon solutions are plane waves of three-momentum $\bm{p}$ with both positive or negative eigenvalues. %$E$. The sign of the energy determines whether we have a particle or an anti-particle; hence this sign can assigned to $u$.     
We interpret an eigenstate with a negative eigenvalue as an anti-particle state with positive energy. %and the sign of the eigenvalue is assigned to the internal quantum number~$u$.  Can't do this. Electrons have u = -2.

The orthonormal sub-basis for the massive spin-$0$ particle is
\begin{align}
	\{\ket{u,0,\bm{p}},
	\text{e}^{\text{i}\theta_\text{PT}}\ket{u,0,-\bm{p}},
	\text{e}^{\text{i}\theta_\text{C}}\ket{-u,0,\bm{p}},\nonumber\\
	\text{e}^{\text{i}\theta_\text{CPT}}\ket{-u,0,-\bm{p}}\}.
\end{align}
Restricting to the sub-group $\{\one,\, CPT\}$,
the representation of the $CPT$ operator in this basis is
\begin{equation}
\label{eq:CPTmatrix}
	CPT
		=\text{e}^{\text{i}\theta_\text{CPT}}
			\begin{pmatrix} 0&0&0&1\\0&0&1&0\\0&1&0&0\\1&0&0&0\end{pmatrix}.
\end{equation}
The resultant $CPT$ operator over this sub-basis is thus unitary and anti-diagonal.
For the massless spin~$0$ particle, the Klein-Gordon equation is simply $\square\psi=0$,
and the space is still spanned by $|\pm u,0,\pm\bm{p}\rangle$
so $CPT$ is the same unitary operator as for the massive-particle case.
Neglecting the unobservable global phase $\theta_\text{CPT}$,
the eigenstates of $CPT$ (\ref{eq:CPTmatrix}) are
\begin{align}\nonumber
	\ket{\pm,0,\bm{p}}\equiv&\frac{1}{\sqrt{2}}\left(\ket{u,0,\bm{p}}\pm\ket{-u,0,-\bm{p}}\right),\\
	\ket{\pm,1,\bm{p}}\equiv&\frac{1}{\sqrt{2}}\left(\ket{u,0,-\bm{p}}\pm\ket{-u,0,\bm{p}}\right)
\label{eq:CPTeigenstates}
\end{align}
with corresponding eigenvalues $\pm 1$. 

We now consider the representation of CPT for a massive Dirac spinor whose state~$\psi$
satisfies the Dirac equation
\begin{equation}
\label{eq:Dirac}
	\left(i\hbar\gamma^\mu\partial_\mu+mc\right)\psi=0.
\end{equation}
The Dirac matrices are
with
\begin{equation}
	\gamma^0=\begin{pmatrix}\one&0\\0&-\one\end{pmatrix},
	\gamma^j=\begin{pmatrix}0&\sigma^j\\-\sigma^j&0\end{pmatrix},
\label{eq9}
\end{equation}
and $\sigma^j|_{j\in(1,2,3)}$ are the Pauli matrices.
Analogous to the previous case of massive $s=0$ particles,
we construct the eight-dimensional state space spanned by $\{\ket{\pm u,\pm1/2,\pm\bm{p}}\}$.
As the action of CPT inverts all degrees of freedom (\ref{eq:CPTstate}), the corresponding unitary $CPT$ matrix is
\begin{equation}
	CPT=\begin{pmatrix}0&0&\sigma^1\\0&\sigma^1&0\\\sigma^1&0&0\end{pmatrix}
\label{eq11}
\end{equation}
with eigenstates 
\begin{align}
		\nonumber
	\ket{\pm,0,\bm{p}}&\equiv\frac{1}{\sqrt{2}}\left(\ket{u,1/2,\bm{p}}\pm\ket{-u,-1/2,-\bm{p}}\right),
		\\ \nonumber
	\ket{\pm,1,\bm{p}}&\equiv\frac{1}{\sqrt{2}}\left(\ket{u,-1/2,\bm{p}}\pm\ket{-u,1/2,-\bm{p}}\right),
		\\ \nonumber
	\ket{\pm,2,\bm{p}}&\equiv\frac{1}{\sqrt{2}}\left(\ket{u,1/2,-\bm{p}}\pm\ket{-u,-1/2,\bm{p}}\right),
		\\
	\ket{\pm,3,\bm{p}}&\equiv\frac{1}{\sqrt{2}}\left(\ket{u,-1/2,-\bm{p}}\pm\ket{-u,1/2,\bm{p}}\right).
\label{eq12}
\end{align}
The eigenvalue for each state
$\{\ket{\pm,\imath,\bm{p}}|_{\imath\in(0,\ldots,3)\}}$ is $\pm 1$.

Dirac spinor states that are invariant under the action of CPT are defined as Majorana spinors~\cite{Duncan:2012fk, G00}. Majorana spinors are also invariant under CPT when obtained as solutions to the Majorana equation
\begin{equation}
\label{eq:Majorana}
	\text{i}\hbar\gamma^{\mu}\partial_{\mu}\psi_\text{c} + mc\psi = 0,
\end{equation}
where, in the Majorana basis, $\psi_\text{c}:= \text{i}\psi^*$. Hence, $\{\one, CPT\}$ is a projective unitary representation of $\mathbb{Z}_2$ for both Dirac and Majorana spinor states.

Massless $s=\frac{1}{2}$ particles are described by the Weyl equation
\begin{equation}
\label{eq:Weyl}
	\text{i}\hbar\sigma^{\mu}\partial_{\mu}\psi = 0
\end{equation}
where the $\sigma^{\mu}$ are the usual Pauli matrices for $\mu=j\in\{1,2,3\}$ and $\sigma^0:=\one$.
The solutions to the Weyl equation~\eqref{eq:Weyl}, known as Weyl spinors, can be represented as four-component spinors.
For $m\equiv 0$, 
Eq.~\eqref{eq:Weyl} is identical to the Dirac equation~\cite{G00}
in which case, the solutions are identical to those of the Dirac equation with states of $\pm u$ representing right-handed and left-handed spinors respectively.
Therefore, the $CPT$ operator is the same for both massive and massless $s=\frac{1}{2}$ particles. 

For massive spin-$1$ particles,
we use the Weinberg-Shay-Good (WSG) equation~\cite{G00}
\begin{equation}
	\left[\text{i}\hbar\partial_{\mu}(\gamma^{\mu\nu}-g^{\mu\nu})i\hbar\partial_{\nu}
		+2m^{2}_{0}c^{2}\right]\psi=0
\label{eq15}
\end{equation}
with $\gamma^{\mu\nu}$ the $6\times 6$ matrices
\begin{align}
	\gamma^{ij=ji}
		=&\begin{pmatrix}0 & \delta_{ij}\one+M^{(ij)}+M^{(ji)} \\ \delta_{ij}\one+M^{(ij)}+M^{(ji)} & 0
		\end{pmatrix}, \nonumber\\
	\gamma^{0i}=&\gamma^{i0}
		=\begin{pmatrix} 0 & S^{i} \\ -S^{i} & 0 \end{pmatrix}, \;
	\gamma^{00}=-\begin{pmatrix} 0 & \one \\ \one & 0 \end{pmatrix},
\label{eq16}
\end{align}
and
\begin{align}
	S^1=&\text{i}\begin{pmatrix}
		0 & 0 & 0 \\0 & 0 & -1 \\0 & 1 & 0
		\end{pmatrix},\;
	S^2=\text{i}\begin{pmatrix}
		0 & 0 & 1 \\0 & 0 & 0 \\-1 & 0 & 0
		\end{pmatrix},
			\nonumber\\
	S^3=&\text{i}\begin{pmatrix}
		0&-1 & 0 \\1 & 0 & 0 \\0 & 0 & 0
		\end{pmatrix},\;
	M^{(ij)}=\text{i}S^j\text{i}S^i.
\label{eq17}
\end{align}
We construct a twelve-dimensional state space spanned by the orthonormal basis
$\{\ket{\pm u,\pm s,\pm\bm{p}}, \, s\in(-1,0,1)\}$.  
The $CPT$ operator is then given by a $12\times 12$ anti-diagonal matrix with unit entries, and the eigenvectors are
\begin{align}
	\ket{\pm,0,\bm{p}}=&\frac{1}{\sqrt{2}}(\ket{u,1,\bm{p}} \pm \ket{-u,-1,-\bm{p}}),
		\nonumber\\
	\ket{\pm,1,\bm{p}}=&\frac{1}{\sqrt{2}}(\ket{u,0,\bm{p}} \pm \ket{-u,0,-\bm{p}}),
		\nonumber\\
	\ket{\pm,2,\bm{p}}=&\frac{1}{\sqrt{2}}(\ket{u,-1,\bm{p}} \pm \ket{-u,1,-\bm{p}}),
		\nonumber\\
	\ket{\pm,3,\bm{p}}=&\frac{1}{\sqrt{2}}(\ket{u,1,-\bm{p}} \pm \ket{-u,-1,\bm{p}}),
		\nonumber\\
	\ket{\pm,4,\bm{p}}=&\frac{1}{\sqrt{2}}(\ket{u,0,-\bm{p}} \pm \ket{-u,0,\bm{p}}),
		\nonumber\\
	\ket{\pm,5,\bm{p}}=&\frac{1}{\sqrt{2}}(\ket{u,-1,-\bm{p}} \pm \ket{-u,1,\bm{p}}), 
\label{eq20}
\end{align}
with eigenvalues $\pm 1$.  Hence, $\{\one,\, CPT\}$ forms a projective unitary representation of $\mathbb{Z}_2$ for massive spin-1 particles.

Now we consider massless $s=1$ particles.
Photons are the only known particles of this type.
The corresponding expression for wave function dynamics
is the Bia{\l}ynicki-Birula--Sipe (BB-S) equation~\cite{Bialynicki-Birula:1994fk,Bialynicki-Birula:1995fk,Bialynicki-Birula:1996uq,Sipe:1995uq,Kobe:1999ve,Smith:2007kx,Raymer:2008zr}
\begin{equation}
	i\hbar\left(\partial_{0}+c\beta^3S^j\partial_{j}\right)\psi = 0,\;
	\beta^3=\begin{pmatrix}\one &0\\0&-\one\end{pmatrix}
\label{eq21}
\end{equation}
with $\{S^{j}\}$ given by Eq.~\eqref{eq17}.
The solutions are six-component spinors with Weyl represention $\psi=\begin{pmatrix}\Psi_+\\ \Psi_-\end{pmatrix}$ where~$\Psi_\pm$ represent opposite helicities~\cite{Smith:2007kx}.

For the solutions of Eq.~\eqref{eq21} to describe a photon correctly,
we require the auxiliary condition~\cite{Bialynicki-Birula:1994fk}
\begin{equation}
	\psi = \beta^1\psi^*,\;
	\beta^1=\begin{pmatrix}0 & \one \\ \one & 0\end{pmatrix}.
\end{equation}
The state space of solutions of Eq.~\eqref{eq21} is the same as for Eq.~\eqref{eq:Weyl}.  Consequently the $CPT$ operator has the same form as that for the massive $s=1$ particles and thus the same eigenvectors. Physically these eigenvectors correspond to linear superpositions of states with opposite helicities.
Such states are usually assumed not to mix and are often treated separately~\cite{Bialynicki-Birula:1994fk,Smith:2007kx}. We also note that the photon does not possess a state of zero total spin (corresponding to a lack of longitudinal polarization in classical optics) and thus the states $\ket{\pm, 1,\bm{p}}$ and $\ket{\pm,4,\bm{p}}$ in Eq.~\eqref{eq20} are unphysical.

For particles of arbitrarily higher spin, solutions may be constructed using the Bargmann-Wigner equations~\cite{BW48}, which are individually indexed Dirac equations.
For instance, $s=\frac{3}{2}$ particles obey a version of the Bargmann-Wigner equations known as the Rarita-Schwinger equation~\cite{RS41} whose solutions are sixteen-component spinors or equivalently four four-component spinors, each of which is individually a solution of the Dirac equation.

We are now ready to formulate a CPT superselection rule. Due to Schur's lemmas~\cite{C89}, unitary representations of finite groups can be fully reduced into their irreducible components (irreps).  In particular, $\mathbb{Z}_2$ has two one-dimensional irreps given by $\pm$.
As CPT superselection implies that coherent superpositions between eigenstates of the $CPT$ operator cannot be observed~\cite{Gour:2008vn}, the state space $\mathscr{H}$ of any system subject to CPT superselection may conveniently be written as 
\begin{equation}
	\mathscr{H}\cong\bigoplus_{\epsilon\in\{\pm\}}\mathscr{H}^{(\epsilon)},
\label{eq:directsum}
\end{equation}
with irrep label $\epsilon$ denoting the two inequivalent irreps of $\mathbb{Z}_2$
and $\mathscr{H}^{(\epsilon)}$ the corresponding eigenspaces.

The Hilbert space~$\mathscr{H}$ in expression~(\ref{eq:directsum})
is applicable only for sub-bases corresponding to a fixed label~$\bm{p}$.
In other words, expression~(\ref{eq:directsum})
is replaced by $\Phi^*_{\bm{p}}\cong\bigoplus_{\epsilon\in\{\pm\}}\Phi_{\bm{p}}^{(\epsilon)}$
with $\Phi^*$ replacing $\mathscr{H}$ because including~$\bm{p}$ means the states
are now distributions, i.e., $\{|u,s,\bm{p}\rangle\}$.
The full space of states corresponds to the space of distributions~$\Phi^*$,
which is a (continuous) sum of all $\Phi^*_{\bm{p}}$.
Partitioning by irrep label~$\epsilon\in\{\pm\}$ holds in the full~$\Phi^*$ as well,
thereby yielding the space of distributions being partitioned into $\Phi^{*(+)}$ and $\Phi^{*(-)}$.
These $\pm$ eigenspaces are spanned by the $CPT$ eigenvectors with positive and negative eigenvalues respectively. 

The states that can be prepared in the absence of a CPT frame of reference (CPT-invariant states) are test  functions that belong in either $\Phi^{(+)}$ or $\Phi^{(-)}$,
which are dual to the spaces $\Phi^{*(+)}$ or $\Phi^{*(-)}$.
Hence, a linear superposition of eigenstates of $CPT$ is a resource and can be brought, via CPT invariant operations, to the standard form
\begin{equation}
	\ket{\psi}=\sqrt{q_0}\ket{+}+\sqrt{q_1}\ket{-},\,
	q_0\in [0,1],\,
	q_1=1-q_0,
\label{eq:standard}
\end{equation}
with $\ket{\pm}$ arbitrary states from $\Phi^{*(\pm)}$.
The important point is that state~(\ref{eq:standard}) is a superposition of two states
chosen from two $\mathbb{Z}_2$ irrep labels~$\pm$.
For simplicity we can consider the state has being in a fixed momentum state,
i.e., a plane wave.
As a perfect plane wave is unphysical, a more realistic treatment would have the state
prepared in a wavepacket with support over a continuum of momentum values~$\bm{p}$.

The frameness inherent in Alice's CPT reference frame token $\ket{\psi}$
is quantified by the alignment rate $R(\psi)$.
This alignment rate quantifies the amount of information Bob obtains 
on average from each reference frame token  $\ket{\psi}$
in the limit of asymptotically many copies~\cite{SG12}. 
For the unitary representations of $\mathbb{Z}_2$~\cite{SG12},
\begin{equation}
	R(\psi)=-2\log\left|q_0-q_1\right|.
\label{eq28}
\end{equation}
If $q_0=q_1=1/2$ in~\eqref{eq:standard} the alignment rate is effectively infinite.

Our strategy for constructing a projective unitary representation of the $CPT$ operator allows for Alice and Bob to communicate information even in the presence of CPT superselection. 
Consider the case that Alice and Bob possess spinless particles.
Alice prepares the plane-wave state 
\begin{equation}
\label{eq:DFS}
	\ket{\phi}=\alpha\ket{+,0,\bm p}+\beta\ket{+,1,\bm p}. 
\end{equation}
As the state~\eqref{eq:DFS} is an eigenstate of the $CPT$ operator for all $\alpha,\,\beta\in\mathbb{C}$, Bob's state is represented exactly the same as Alice's after correcting for the known
Poincar\'{e} transformation~$\Lambda$ between their reference frames.
By choosing the coefficients $\alpha, \,\beta$ appropriately,
Alice can encode a logical qubit,
which Bob can retrieve by performing the appropriate decoding. 

Here we have shown that the superselection rule arising from CPT symmetry can be circumvented using CPT frameness resources. We have identified the ultimate frameness resources for the cases of both massive and massless spin~$0$, $\frac{1}{2}$, and~$1$ particles including Majorana spinors and have suggested a strategy for finding solutions for states of arbitrary spin. We have also shown that communication is possible, even in the presence of CPT superselection, except for the case of spinless particles with three-momentum equal to zero. 

MS acknowledges financial support from NSERC, USARO, and Austrian science fund (FWF) Y535-N16, P20748-N16, P24273-N16 and F40-FoQus F4011/12-N16. ITD acknowledges financial support from FQXi.  BCS acknowledges support from NSERC, AITF and CIFAR. 
 
\bibliographystyle{apsrev}
\bibliography{cpt.bib}

\begin{thebibliography}{30}
\expandafter\ifx\csname natexlab\endcsname\relax\def\natexlab#1{#1}\fi
\expandafter\ifx\csname bibnamefont\endcsname\relax
  \def\bibnamefont#1{#1}\fi
\expandafter\ifx\csname bibfnamefont\endcsname\relax
  \def\bibfnamefont#1{#1}\fi
\expandafter\ifx\csname citenamefont\endcsname\relax
  \def\citenamefont#1{#1}\fi
\expandafter\ifx\csname url\endcsname\relax
  \def\url#1{\texttt{#1}}\fi
\expandafter\ifx\csname urlprefix\endcsname\relax\def\urlprefix{URL }\fi
\providecommand{\bibinfo}[2]{#2}
\providecommand{\eprint}[2][]{\url{#2}}

\bibitem[{\citenamefont{Wick et~al.}(1952)\citenamefont{Wick, Wightman, and
  Wigner}}]{Wick:1952ys}
\bibinfo{author}{\bibfnamefont{G.~C.} \bibnamefont{Wick}},
  \bibinfo{author}{\bibfnamefont{A.~S.} \bibnamefont{Wightman}},
  \bibnamefont{and} \bibinfo{author}{\bibfnamefont{E.~P.}
  \bibnamefont{Wigner}}, \bibinfo{journal}{Phys. Rev.}
  \textbf{\bibinfo{volume}{88}}, \bibinfo{pages}{101} (\bibinfo{year}{1952}).

\bibitem[{\citenamefont{Aharonov and Susskind}(1967)}]{Aharonov:1967zr}
\bibinfo{author}{\bibfnamefont{Y.}~\bibnamefont{Aharonov}} \bibnamefont{and}
  \bibinfo{author}{\bibfnamefont{L.}~\bibnamefont{Susskind}},
  \bibinfo{journal}{Phys. Rev.} \textbf{\bibinfo{volume}{155}},
  \bibinfo{pages}{1428} (\bibinfo{year}{1967}).

\bibitem[{\citenamefont{Bartlett et~al.}(2003)\citenamefont{Bartlett, Rudolph,
  and Spekkens}}]{BRS03}
\bibinfo{author}{\bibfnamefont{S.~D.} \bibnamefont{Bartlett}},
  \bibinfo{author}{\bibfnamefont{T.}~\bibnamefont{Rudolph}}, \bibnamefont{and}
  \bibinfo{author}{\bibfnamefont{R.~W.} \bibnamefont{Spekkens}},
  \bibinfo{journal}{Phys. Rev. Lett.} \textbf{\bibinfo{volume}{91}},
  \bibinfo{pages}{027901} (\bibinfo{year}{2003}).

\bibitem[{\citenamefont{Collins et~al.}(2005)\citenamefont{Collins, Diosi,
  Gisin, Massar, and Popescu}}]{Collins:2005uq}
\bibinfo{author}{\bibfnamefont{D.}~\bibnamefont{Collins}},
  \bibinfo{author}{\bibfnamefont{L.}~\bibnamefont{Diosi}},
  \bibinfo{author}{\bibfnamefont{N.}~\bibnamefont{Gisin}},
  \bibinfo{author}{\bibfnamefont{S.}~\bibnamefont{Massar}}, \bibnamefont{and}
  \bibinfo{author}{\bibfnamefont{S.}~\bibnamefont{Popescu}},
  \bibinfo{journal}{Phys. Rev. A} \textbf{\bibinfo{volume}{72}},
  \bibinfo{pages}{022304} (\bibinfo{year}{2005}).

\bibitem[{\citenamefont{Gisin}(2004)}]{Gisin:2004fk}
\bibinfo{author}{\bibfnamefont{N.}~\bibnamefont{Gisin}},
  \bibinfo{journal}{arXiv:quant-ph/0408095}  (\bibinfo{year}{2004}).

\bibitem[{\citenamefont{Rudolph and Sanders}(2001)}]{Rudolph:2001uq}
\bibinfo{author}{\bibfnamefont{T.}~\bibnamefont{Rudolph}} \bibnamefont{and}
  \bibinfo{author}{\bibfnamefont{B.~C.} \bibnamefont{Sanders}},
  \bibinfo{journal}{Phys. Rev. Lett.} \textbf{\bibinfo{volume}{87}},
  \bibinfo{pages}{077903} (\bibinfo{year}{2001}).

\bibitem[{\citenamefont{Sanders et~al.}(2003)\citenamefont{Sanders, Bartlett,
  Rudolph, and Knight}}]{Sanders:2003kx}
\bibinfo{author}{\bibfnamefont{B.~C.} \bibnamefont{Sanders}},
  \bibinfo{author}{\bibfnamefont{S.~D.} \bibnamefont{Bartlett}},
  \bibinfo{author}{\bibfnamefont{T.}~\bibnamefont{Rudolph}}, \bibnamefont{and}
  \bibinfo{author}{\bibfnamefont{P.~L.} \bibnamefont{Knight}},
  \bibinfo{journal}{Phys. Rev. A} \textbf{\bibinfo{volume}{68}},
  \bibinfo{pages}{042329} (\bibinfo{year}{2003}).

\bibitem[{\citenamefont{Bartlett et~al.}(2007)\citenamefont{Bartlett, Rudolph,
  and Spekkens}}]{Bartlett:2007fk}
\bibinfo{author}{\bibfnamefont{S.~D.} \bibnamefont{Bartlett}},
  \bibinfo{author}{\bibfnamefont{T.}~\bibnamefont{Rudolph}}, \bibnamefont{and}
  \bibinfo{author}{\bibfnamefont{R.~W.} \bibnamefont{Spekkens}},
  \bibinfo{journal}{Rev. Mod. Phys.} \textbf{\bibinfo{volume}{79}},
  \bibinfo{pages}{555} (\bibinfo{year}{2007}).

\bibitem[{\citenamefont{Gour and Spekkens}(2008)}]{Gour:2008vn}
\bibinfo{author}{\bibfnamefont{G.}~\bibnamefont{Gour}} \bibnamefont{and}
  \bibinfo{author}{\bibfnamefont{R.~W.} \bibnamefont{Spekkens}},
  \bibinfo{journal}{New J. Phys.} \textbf{\bibinfo{volume}{10}},
  \bibinfo{pages}{033023} (\bibinfo{year}{2008}).

\bibitem[{\citenamefont{Schwinger}(1951)}]{Schwinger:1951kx}
\bibinfo{author}{\bibfnamefont{J.}~\bibnamefont{Schwinger}},
  \bibinfo{journal}{Phys. Rev.} \textbf{\bibinfo{volume}{82}},
  \bibinfo{pages}{914} (\bibinfo{year}{1951}).

\bibitem[{\citenamefont{L\"{u}ders}(1954)}]{Luders:1954fk}
\bibinfo{author}{\bibfnamefont{G.}~\bibnamefont{L\"{u}ders}},
  \bibinfo{journal}{Kongelige Danske Vidensabernes Selskab Matematisk-fysiske
  Meddelelser} \textbf{\bibinfo{volume}{28}}, \bibinfo{pages}{1}
  (\bibinfo{year}{1954}).

\bibitem[{\citenamefont{Pauli}(1955)}]{Pauli:1955uq}
\bibinfo{author}{\bibfnamefont{W.}~\bibnamefont{Pauli}},
  \emph{\bibinfo{title}{Niels Bohr and the Development of Physics}}
  (\bibinfo{publisher}{Pergamon Press}, \bibinfo{address}{New York},
  \bibinfo{year}{1955}).

\bibitem[{\citenamefont{Scurek}(2004)}]{Scu04}
\bibinfo{author}{\bibfnamefont{R.}~\bibnamefont{Scurek}},
  \bibinfo{journal}{American Journal of Physics} \textbf{\bibinfo{volume}{72}},
  \bibinfo{pages}{638} (\bibinfo{year}{2004}),
  \urlprefix\url{http://link.aip.org/link/?AJP/72/638/1}.

\bibitem[{\citenamefont{Gour et~al.}(2009)\citenamefont{Gour, Sanders, and
  Turner}}]{Gour:2009ly}
\bibinfo{author}{\bibfnamefont{G.}~\bibnamefont{Gour}},
  \bibinfo{author}{\bibfnamefont{B.~C.} \bibnamefont{Sanders}},
  \bibnamefont{and} \bibinfo{author}{\bibfnamefont{P.~S.}
  \bibnamefont{Turner}}, \bibinfo{journal}{J. Math. Phys.}
  \textbf{\bibinfo{volume}{50}}, \bibinfo{pages}{102105}
  (\bibinfo{year}{2009}).

\bibitem[{\citenamefont{Feynman}(1949)}]{Feynman:1949uq}
\bibinfo{author}{\bibfnamefont{R.~P.} \bibnamefont{Feynman}},
  \bibinfo{journal}{Phys. Rev.} \textbf{\bibinfo{volume}{76}},
  \bibinfo{pages}{749} (\bibinfo{year}{1949}).

\bibitem[{\citenamefont{Bell and Jackiw}(1969)}]{Bell:1969uq}
\bibinfo{author}{\bibfnamefont{J.~S.} \bibnamefont{Bell}} \bibnamefont{and}
  \bibinfo{author}{\bibfnamefont{R.}~\bibnamefont{Jackiw}},
  \bibinfo{journal}{Il Nuovo Cimento A} \textbf{\bibinfo{volume}{60}},
  \bibinfo{pages}{47} (\bibinfo{year}{1969}).

\bibitem[{\citenamefont{Gel'fand and Vilenkin}(1964)}]{GelfandVilenkin64}
\bibinfo{author}{\bibfnamefont{I.~M.} \bibnamefont{Gel'fand}} \bibnamefont{and}
  \bibinfo{author}{\bibfnamefont{N.~Y.} \bibnamefont{Vilenkin}},
  \emph{\bibinfo{title}{Generalized Functions}} (\bibinfo{publisher}{Academic,
  New York}, \bibinfo{year}{1964}).

\bibitem[{\citenamefont{Greiner}(2000)}]{G00}
\bibinfo{author}{\bibfnamefont{W.}~\bibnamefont{Greiner}},
  \emph{\bibinfo{title}{Relativistic {Q}uantum {M}echanics {W}ave {E}quations,
  Third edition}} (\bibinfo{publisher}{Springer-Verlag, Heidelberg},
  \bibinfo{year}{2000}).

\bibitem[{\citenamefont{Bargmann and Wigner}(1948)}]{BW48}
\bibinfo{author}{\bibfnamefont{V.}~\bibnamefont{Bargmann}} \bibnamefont{and}
  \bibinfo{author}{\bibfnamefont{E.~P.} \bibnamefont{Wigner}},
  \bibinfo{journal}{Proc. Nat. Acad. Sci.} \textbf{\bibinfo{volume}{34}},
  \bibinfo{pages}{211} (\bibinfo{year}{1948}).

\bibitem[{\citenamefont{Duncan}(2012)}]{Duncan:2012fk}
\bibinfo{author}{\bibfnamefont{A.}~\bibnamefont{Duncan}},
  \emph{\bibinfo{title}{The Conceptual Framework of Quantum Field Theory}}
  (\bibinfo{publisher}{Oxford University Press}, \bibinfo{address}{Oxford},
  \bibinfo{year}{2012}).

\bibitem[{\citenamefont{Bialynicki-Birula}(1994)}]{Bialynicki-Birula:1994fk}
\bibinfo{author}{\bibfnamefont{I.}~\bibnamefont{Bialynicki-Birula}},
  \bibinfo{journal}{Acta Physica Polonica A} \textbf{\bibinfo{volume}{86}},
  \bibinfo{pages}{97} (\bibinfo{year}{1994}).

\bibitem[{\citenamefont{Bialynicki-Birula}(1995)}]{Bialynicki-Birula:1995fk}
\bibinfo{author}{\bibfnamefont{I.}~\bibnamefont{Bialynicki-Birula}}, in
  \emph{\bibinfo{booktitle}{Coherence and Quantum Optics VII}}, edited by
  \bibinfo{editor}{\bibfnamefont{J.}~\bibnamefont{Eberly}},
  \bibinfo{editor}{\bibfnamefont{L.}~\bibnamefont{Mandel}}, \bibnamefont{and}
  \bibinfo{editor}{\bibfnamefont{E.}~\bibnamefont{Wolf}}
  (\bibinfo{publisher}{Plenum}, \bibinfo{address}{New York},
  \bibinfo{year}{1995}), p. \bibinfo{pages}{313}.

\bibitem[{\citenamefont{Bialynicki-Birula}(1996)}]{Bialynicki-Birula:1996uq}
\bibinfo{author}{\bibfnamefont{I.}~\bibnamefont{Bialynicki-Birula}}, in
  \emph{\bibinfo{booktitle}{Progress in Optics XXXVI}}, edited by
  \bibinfo{editor}{\bibfnamefont{E.}~\bibnamefont{Wolf}}
  (\bibinfo{publisher}{Elsevier}, \bibinfo{address}{Amsterdam},
  \bibinfo{year}{1996}), p. \bibinfo{pages}{245}.

\bibitem[{\citenamefont{Sipe}(1995)}]{Sipe:1995uq}
\bibinfo{author}{\bibfnamefont{J.~E.} \bibnamefont{Sipe}},
  \bibinfo{journal}{Phys. Rev. A} \textbf{\bibinfo{volume}{52}},
  \bibinfo{pages}{1875} (\bibinfo{year}{1995}).

\bibitem[{\citenamefont{Kobe}(1999)}]{Kobe:1999ve}
\bibinfo{author}{\bibfnamefont{D.}~\bibnamefont{Kobe}},
  \bibinfo{journal}{Foundations of Physics} \textbf{\bibinfo{volume}{29}},
  \bibinfo{pages}{1203} (\bibinfo{year}{1999}).

\bibitem[{\citenamefont{Smith and Raymer}(2007)}]{Smith:2007kx}
\bibinfo{author}{\bibfnamefont{B.~J.} \bibnamefont{Smith}} \bibnamefont{and}
  \bibinfo{author}{\bibfnamefont{M.}~\bibnamefont{Raymer}},
  \bibinfo{journal}{New Journal of Physics} \textbf{\bibinfo{volume}{9}},
  \bibinfo{pages}{414} (\bibinfo{year}{2007}).

\bibitem[{\citenamefont{Raymer and Smith}(2008)}]{Raymer:2008zr}
\bibinfo{author}{\bibfnamefont{M.}~\bibnamefont{Raymer}} \bibnamefont{and}
  \bibinfo{author}{\bibfnamefont{B.~J.} \bibnamefont{Smith}}, in
  \emph{\bibinfo{booktitle}{The Nature of Light: What is a Photon?}}, edited by
  \bibinfo{editor}{\bibfnamefont{C.}~\bibnamefont{Roychoudhuri}},
  \bibinfo{editor}{\bibfnamefont{A.}~\bibnamefont{Kracklauer}},
  \bibnamefont{and} \bibinfo{editor}{\bibfnamefont{K.}~\bibnamefont{Creath}}
  (\bibinfo{publisher}{CRC Press}, \bibinfo{address}{Boca Raton},
  \bibinfo{year}{2008}), pp. \bibinfo{pages}{207--214}.

\bibitem[{\citenamefont{Rarita and Schwinger}(1941)}]{RS41}
\bibinfo{author}{\bibfnamefont{W.}~\bibnamefont{Rarita}} \bibnamefont{and}
  \bibinfo{author}{\bibfnamefont{J.}~\bibnamefont{Schwinger}},
  \bibinfo{journal}{Phys. Rev.} \textbf{\bibinfo{volume}{60}},
  \bibinfo{pages}{61} (\bibinfo{year}{1941}),
  \urlprefix\url{http://link.aps.org/doi/10.1103/PhysRev.60.61}.

\bibitem[{\citenamefont{Chen}(1989)}]{C89}
\bibinfo{author}{\bibfnamefont{J.-Q.} \bibnamefont{Chen}},
  \emph{\bibinfo{title}{Group {R}epresentation {T}heory for {P}hysicists}}
  (\bibinfo{publisher}{World Scientific, Singapore}, \bibinfo{year}{1989}).

\bibitem[{\citenamefont{Skotiniotis and Gour}(2012)}]{SG12}
\bibinfo{author}{\bibfnamefont{M.}~\bibnamefont{Skotiniotis}} \bibnamefont{and}
  \bibinfo{author}{\bibfnamefont{G.}~\bibnamefont{Gour}}, \bibinfo{journal}{New
  J. Phys.} \textbf{\bibinfo{volume}{14}}, \bibinfo{pages}{073022}
  (\bibinfo{year}{2012}).

\end{thebibliography}
\end{document}